\documentstyle[prd,epsf,aps]{revtex}
\newcommand{\ds}{\displaystyle}

\newcommand{\nn}{\nonumber\\}
\newcommand{\be}{\begin{equation}}
\newcommand{\ee}{\end{equation}}
\newcommand{\ba}{\begin{eqnarray}}
\newcommand{\ea}{\end{eqnarray}}
\newcommand{\by}{\begin{eqnarray*}}
\newcommand{\ey}{\end{eqnarray*}}

\newcommand{\La}{\Lambda}

\newcommand{\Om}{\Omega}
\newcommand{\om}{\omega}

\newcommand{\Del}{\Delta}

\newcommand{\si}{\sigma}

\newcommand{\f}{\frac}
\newcommand{\ti}{\tilde}

\newcommand\mvec[1]{{\mathbf{.)1}}}
\newcommand\vep\varepsilon

\newcommand\up[1]{{\rm#1}}

\title{ \bf Chromomagnetic Catalysis of Chiral Symmetry Breaking and
Color Superconductivity \footnote{Invited talk given by one of us
(D.E.) at the workshop "Hadron structure and reactions from 
nonperturbative QCD" at RCNP, Osaka University, Japan, July 23-25,
2001.}}
\author{D.~Ebert $^{\flat,\dagger}$, V.V. Khudyakov $^{\star}$,
K.~G.~Klimenko $^{\ddagger}$, H.~
Toki $^{\flat}$, and V.~Ch.~Zhukovsky $^{\star}$}
\address{$^{\flat}$ Research
Center for Nuclear Physics (RCNP), Osaka University, Ibaraki,Osaka
567,Japan}
\address{$^{\dagger}$ Institut f\"ur Physik,
Humboldt-Universit\"at zu Berlin, 10115 Berlin, Germany}
\address{$^{\star}$ Faculty of
Physics, Department of Theoretical Physics, Moscow State University,
119899, Moscow, Russia}
\address{$^{\ddagger}$ Institute
of High Energy Physics, 142284, Protvino, Moscow Region, Russia}
\begin{document}
\large
\maketitle
\begin{abstract}
It is shown in the framework of an extended NJL model with two
flavors
that some types of external chromomagnetic field induce the dynamical
chiral or color symmetry breaking even at weakest attraction between
quarks. It is argued also that an external chromomagnetic field,
simulating the 
chromomagnetic gluon condensate of the real QCD-vacuum, might
significantly 
influence the color superconductivity formation.
\end{abstract}
\renewcommand{\thefootnote}{\arabic{footnote}}
\setcounter{footnote}{0}
\setcounter{page}{1}
\section{ Introduction}

At the beginning of the last decade an exciting property of the
homogeneous external magnetic field to dynamically generate the
chiral symmetry breaking (CSB) even at the weakest attractive forces
between fermions has been discovered \cite{k1}-\cite{gus2}. Now it is
well-known as magnetic catalysis effect (MCE). 

First particular observations of MCE were done in \cite{k1} on 
the basis of a (2+1)-dimensional model with four fermion 
interaction. Then, it was shown that in 3D this
effect is a model independent one, and the explanation for
MCE was given in the framework of a dimensional reduction mechanism
\cite{gus1}. The investigation of MCE under the influence of
different
external factors and for the case of (3+1)-dimensional models are
given in \cite{mir}. Besides, this phenomenon finds interesting 
applications in cosmology \cite{10} and condensed matter physics
\cite{liu} (see also the reviews \cite{gus2} and references therein).

Later a similar property of the
homogeneous external chromomagnetic field to dynamically generate the
CSB has been found as well \cite{k2}-\cite{zhuk}.
The physical essence of this effect is again the effective reduction
of the space-time dimensionality in the presence of external 
chromomagnetic fields \cite{zhuk}. 
Recently, it was also shown in the framework of a
Nambu -- Jona-Lasinio (NJL) model that some types of chromomagnetic
fields might induce the dynamical breaking of the color symmetry,
thus
catalizing the appearance of color superconductivity
\cite{e1,1,2}.

In accordance with modern knowledge, the QCD vacuum at low
temperature and density is characterized by the confinement
phemomenon, i.e. quarks and gluons are not observed, since they are
confined into hadrons, and the color symmetry is not broken. 
Two nonperturbative features are inherent to the QCD vacuum in this
phase. One is the nonzero value of the gluon condensate $\langle
FF\rangle\equiv$$\langle F^a_{\mu\nu}F^{a\mu\nu}\rangle$,
where $F^a_{\mu\nu}$ is the field strength tensor of the gluon
fields.
Another one is
the nonzero chiral condensate $\langle\bar qq\rangle$
which signals about CSB. At high temperatures the quark-gluon plasma
phase is expected to exist. In this phase all symmetries of the QCD
Lagrangian are restored, and quarks and gluons are elementary
excitations of the theory.
 It was realized more than twenty years ago \cite{Ba},
that at high densities (high values of the chemical potential $\mu$) 
the color superconducting (CSC) phase might exist. The CSC-vacuum
is generated by the condensation of quark Cooper pairs, i.e.
the vacuum expectation value of diquarks $\langle qq\rangle$ is
nonzero. Since quark Cooper pairing occurs in
the color
anti-triplet channel, the nonzero value of $\langle qq\rangle$ means
that, apart from the electromagnetic $U(1)$ symmetry,  the color
$SU_c(3)$ should be spontaneously broken down  inside
the CSC phase as well. 

The CSC phenomenon was investigated in the framework of the one-gluon
exchange approximation in QCD \cite{son}, where
the colored Cooper pair formation is predicted selfconsistently
at extremely high values of the chemical potential $\mu\gtrsim 10^8$
MeV \cite{raj}. Unfortunately, such baryon densities are not
observable
in nature and not accessible in experiments (the typical
densities inside the neutron stars or in the future heavy ion
experiments correspond to  $\mu\sim 500$ MeV). The
possibility for the existency of the CSC phase in the region
of moderate densities was proved quite recently
(see e.g. the papers \cite{rapp}-\cite{neb} 
as well as the review article \cite{alf2} and references therein).
In these papers it was shown on the basis of different effective
theories for low energy QCD (instanton model, NJL model etc)
that the diquark condensate $\langle qq\rangle$ can appear  already
at a rather moderate baryon density ($\mu\sim 400$ MeV), which can
possibly be detected in the future experiments on heavy ion-ion
collisions.

In the framework of NJL models the CSC phase formation has generally 
be considered as a dynamical competition between diquark $\langle
qq\rangle$ and usual quark-antiquark condensation $\langle\bar
qq\rangle$. However, the real QCD vacuum is characterized in addition
by the appearence of a gluon condensate $\langle FF\rangle$ as well,
which might change the generally accepted conditions for the CSC
observation. As an effective theory for low energy QCD, the NJL model
does not contain any dynamical gluon fields. As a consequence, the
nonzero value of $\langle FF\rangle$ cannot be generated
dynamically in this scheme, but it can be
mimicked with the help of external chromomagnetic fields. 
In particular, for a 
QCD-motivated NJL model with gluon condensate (i.e. in the 
presence of an external chromomagnetic field) and finite
temperature, it was shown that a weak gluon condensate plays a
stabilizing role for the behavior of the constituent quark mass,
the quark condensate, meson masses and coupling constants for
varying temperature \cite{8}. 

The aim of the present talk is to discuss the influence of external
conditions, such as the chemical potential
and especially
the gluon condensate (as modelled by external color gauge fields),
on the phase structure of quark matter with particular emphasize of
its CSC phase. To this end, we have extended our earlier analysis of
the chromomagnetic generation of CSC at $\mu=0$ \cite{e1}
to the case of an (3+1)-dimensional  NJL type model  with finite
chromomagnetic field
and chemical potential presenting a
generalization of the zero external field model of \cite{klev}.

The talk is organized as follows. In Section II the
extended NJL model under consideration is presented, and its
effective potential ($\equiv$ thermodynamic potential) at nonzero
external chromomagnetic field and chemical potential
is presented in the one-loop approximation. This quantity contains
all
the necessary informations about the quark and diquark condensates of
the theory. It is well-known that the chemical potential is a factor, 
which promotes the generation of CSC.
We argue that an external chromomagnetic field is another
factor with similar properties. 
In order to prove this statement
 we first consider in Section III the simpler case with zero 
chemical potential. It is shown here that some types of the
external chromomagnetic field can induce the transitions
to the CSB or CSC phases even at weakest
quark interaction (depending on the relation between
couplings in $\bar qq$ and $qq$ channels). The combined influence of
both chemical potential and external chromomagnetic field
on the generation of  $\langle\bar qq\rangle$ and  $\langle
qq\rangle$  condensates at
physically meaningful values of coupling constants 
is considered in Section IV. It is shown there that 
the characteristics of the CSC phase significantly 
depend on the strength of the chromomagnetic field. 
Finally, section V contains 
a summary and discussion of the results. 

\section{The model and the effective potential}

Let us first give several (very approximative) arguments motivating
the chosen
structure of our QCD-inspired extended NJL model introduced below.
For this aim, consider two-flavor QCD with nonzero chemical 
potential and color group $SU_c(3)$ and decompose the gluon field
${\cal A}_\nu^a(x)$ into a condensate background (``external'') field
$A_\nu^a(x)$ and the quantum fluctuation $a_\nu^a(x)$ around it, i.e. 
${\cal A}_\nu^a(x)=$$A_\nu^a(x)+$$a_\nu^a(x).$
By integrating in the generating functional of QCD over
the quantum field $a_\nu^a(x)$ and further ``approximating'' the 
nonperturbative gluon propagator by a $\delta-$function,
one arrives at an effective local chiral four-quark interaction of
the NJL type describing low energy hadron physics 
in the presence of a gluon condensate. Finally, by performing a
Fierz transformation of the interaction term, one obtains 
a four-fermionic model with $(\bar q q)$--and $(q q)$--interactions
and an external condensate field $A_\mu^a(x)$
of the color group $SU_c(N_c)$  given by the following
Lagrangian
\footnote{The most general four-fermion interaction would
include additional vector and axial-vector $(\bar q q)$ as
well as pseudo-scalar, vector and axial-vector-like $(q q)$
-interactions. For our goal of studying the effect of
chromomagnetic catalysis for the competition of quark and diquark
condensates, the interaction structure of (1) is, however,
sufficiently general.}
\begin{eqnarray}
  L=\bar q[\gamma^\nu(i\partial_\nu+g
 A_\nu^a(x)\frac{\lambda_a}2)+\mu\gamma^0]q
  +\frac{G_1}{2N_c}[(\bar
  qq)^2+(\bar qi\gamma^5\vec
  \tau q)^2]+
  \frac{G_2}{N_c}[i\bar q_c\varepsilon
\epsilon^b \gamma^5 q]
  [i\bar q\varepsilon\epsilon^b\gamma^5 q_c],
  \label{1}
\end{eqnarray}
It is necessary to note that in order to obtain 
realistic estimates for masses of 
vector/axial-vector mesons and diquarks 
in extended NJL--type of models \cite{ebertt}, we have to allow for
independent coupling constants 
$G_1, G_2$, rather than to consider them related by a Fierz
transformation
of a current-current interaction via gluon exchange. Clearly, such a  
procedure does not spoil chiral symmetry. 

In (\ref{1}) $g$ denotes the gluon coupling constant,
$\mu$ is the quark chemical potential, $q_c=C\bar
q^t$, $\bar q_c=q^t C$ are charge-conjugated spinors,
and $C=i\gamma^2\gamma^0$ 
is the charge conjugation matrix ($t$ denotes
the transposition operation). 
In what follows we assume $N_c=3$. 
Moreover, summation over repeated color indices $a =1,\dots,8$; $b =
1,2,3$ and Lorentz indices
$\nu = 0,1,2,3$ is implied. The quark field $q\equiv q_{i\alpha}$ is
a flavor doublet and color triplet as well as a four-component Dirac
spinor, where $i=1,2$; $\alpha = 1,2,3$. (Latin and Greek indices
refer to flavor and color indices, respectively; spinor indices are
omitted.) Furthermore, we use the notations $\lambda^a/2$ for the
generators of the color $SU_c(3)$ group 
appearing in the covariant derivative 
as well as $\vec \tau\equiv (\tau^{1},
\tau^{2},\tau^3)$ for Pauli matrices in the flavor space; 
$(\varepsilon)^{ik}\equiv\varepsilon^{ik}$,
$(\epsilon^b)^{\alpha\beta}\equiv\epsilon^{\alpha\beta b}$
are totally antisymmetric tensors in the flavor and color spaces,
respectively. Clearly, the Lagrangian (\ref{1}) is invariant under
the chiral $SU(2)_L\times SU(2)_R$ and color $SU_c(3)$ groups.

Next, let us for a  moment suppose that in (\ref{1}) 
$A_\mu^a(x)$ is an arbitrary
classical gauge field of the color group $SU_c(3)$. (The following
investigations do not require the explicit inclusion of the gauge
field part of the Lagrangian). The detailed structure of $A_\mu^a(x)$
corresponding to a constant chromomagnetic gluon condensate will be
given below.

The linearized version of the model (\ref{1}) with auxiliary bosonic
fields has the following form
\begin{eqnarray}
\tilde L\ds &=&\bar q[\gamma^\nu(i\partial_\nu+g
 A_\nu^a(x)\frac{\lambda_a}2)+\mu\gamma^0]q
 -\bar q(\sigma+i\gamma^5\vec
 \tau\vec\pi)q-\frac{N_c}{2G_1}(\sigma^2+\vec \pi^2)-
 \nonumber\\
&-&\frac{N_c}{G_2}\Delta^{*b}\Delta^b-
\Delta^{*b}[iq^tC\varepsilon\epsilon^b\gamma^5 q]
 -\Delta^b[i\bar q \varepsilon\epsilon^b\gamma^5 C\bar q^t].
 \label{2}
\end{eqnarray}
The Lagrangians (\ref{1}) and (\ref{2}) are equivalent, as can be
seen by using the equations of motion for bosonic fields, from which
it follows that
\begin{equation}
  \Delta^b\sim iq^t C\varepsilon \epsilon^b\gamma^5 q,\quad
  \sigma\sim\bar qq,\quad
  \vec \pi\sim i\bar q\gamma^5\vec\tau q.
\end{equation}
Clearly, $\sigma$ and $\vec\pi$ fields are color singlets.
Besides, the (bosonic) diquark field $\Delta^{b}$ is a color
antitriplet and a singlet under the chiral
$SU(2)_L\times SU(2)_R$ group. Note further that $\sigma$,
$\Delta^b$, are scalars, but $\vec\pi$ are pseudo-scalar fields.
Hence, if $\langle\sigma\rangle\ne0$, then chiral
symmetry of the model is spontaneously broken, whereas 
$\langle\Delta^b\rangle\ne0$ indicates the dynamical
breaking of both the color and electromagnetic  symmetries of the
theory.

In the one-loop approximation, the effective action for the boson
fields is expressed through the path integral over quark fields:
\be
  \exp(i S_{\up{eff}}(\sigma,\vec\pi,\Delta^b,\Delta^{*b},A_\mu^a))=
  N'\int[d\bar q][dq]\exp\Bigl(i\int\tilde L\,d^4 x\Bigr),
\label{k3}
\ee
where $N'$ is a normalization constant.
Suppose that all the boson fields in  (\ref{k3}), except $A_\mu^a$,
do not depend on the space-time points. Since $S_{\up{eff}}$ is a
function invariant under the chiral (flavor) as well as color and
Lorentz groups, it is possible to find a frame in which
$\Delta^1$=$\Delta^2$=$\vec\pi=0$, i.e.
$S_{\up{eff}}\equiv$$S_{\up{eff}}(\sigma,\Delta)$, where
$\Delta\equiv\Delta^3$. Next, let us define the effective potential
by the following relation
$S_{\up{eff}}(\sigma,\Delta)\equiv$$-V_{\up{eff}}(\sigma,\Delta)\int
d^4x$. The global minimum point of $V_{\rm eff}$ defines
the vacuum expectation values of boson fields as well as the vacuum
residual symmetry group. For example, if in this point
$\Delta\equiv\langle\Delta\rangle\ne 0$, then $SU_c(3)$ is broken
up to $SU_c(2)$, whose generators are the first three generators of
initial $SU_c(3)$, and the CSC phenomenon is observed.
 Using pure symmetry arguments, it is easily
shown that, if dynamical gluons were introduced into consideration,
the three gluons living in the unbroken SU$_c(2)$ subgroup
would stay massless, whereas the remaining five gluons would get
masses.  Correspondingly, in this frame the external
chromomagnetic field $H^a$ can be represented in the following way:
$H^a=H_I^a+H^a_{II}$, where
$H^a_I=(H^1, H^2, H^3,0,\dots,0)$, $H^a_{II}=(0,0,0,H^4,\dots,H^8)$.
By analogy with ordinary superconductivity, it is expected that
external chromomagnetic fields corresponding to massive gluons,
i.e. external chromomagnetic fields of the form $H^a_{II}$,
should be expelled from the CSC phase (Meissner effect).
Moreover, sufficiently high values of such fields should destroy the
CSC.  However, our intuition tells us nothing about the action
of external chromomagnetic fields of the form $H_I^a$ on the color
superconducting state of the quark-gluon system.

In the present talk the influence of such external
chromomagnetic fields of the form $H^a=(H^1,H^2,H^3,0,...,0)$  on the
phase structure of the NJL model is considered. Furthermore,
due to the residual $SU_c(2)$ invariance of the vacuum, 
one can put $H^{1}= H^2=0$ and $H^{3}\equiv H.$
Correspondingly, the gluon condensate which is mimicked by this
external field has the value $\langle FF\rangle=2H^2$.
Next, some remarks about the structure of the external 
vector-potential $A_\nu^a(x)$ used in (\ref{1}) are needed. From this
moment on, we assume $A_\nu^a(x)$ in such a form that the only
nonvanishing components of the
corresponding field strength tensor $F_{\mu\nu}^a$ are
$F_{12}^3=-F_{21}^3=H=\up{const}$.
The above homogeneous chromomagnetic field can be generated by the
following vector-potential
\begin{equation}
  A_\nu^3(x)=(0,0,Hx^1,0);\quad A_\nu^a(x)=0 \quad (a\ne 3),
  \label{x7}
\end{equation}
which defines the well known Matinyan--Savvidy model of the gluon
condensate in QCD \cite{savvidy}.
In $QCD$ the physical vacuum may be interpreted as a region splitted
into an infinite number of domains with macroscopic extension
\cite{nielsen}. Inside each such domain there can be excited a
homogeneous background chromomagnetic field, which generates a
nonzero gluon condensate $\langle FF\rangle$.  (Averaging over
all domains results in a zero background chromomagnetic field, hence
color as well as Lorentz symmetries are not broken.)
Recall, that in order to find condensates
$\langle\sigma\rangle$ and
$\langle\Delta\rangle$, we should calculate the effective potential
whose global minimum point provides us with these quantities.
The expression for the effective potential at $\mu\ne 0$, $H\ne 0$,
$T=0$ has the following form \cite{e1}-\cite{2}:
\begin{eqnarray}
\label{eq.8}
V_{H\mu}(\sigma,\Delta)=\frac{3\sigma^2}{2G_1}+\frac
{3\Delta\Delta^{\ast}}{G_2}-\frac{\tilde
S(\sigma,\Delta)}{v},~~v=\int d^4x,
\end{eqnarray}
where
\begin{eqnarray}
\exp (i\tilde S(\sigma,\Delta))&=&N'{\rm
det}[(i\hat\partial-\sigma+\mu\gamma^0)]\cdot\nonumber\\
\cdot{\rm det}^{1/2}\left [4|\Delta|^2
\right.&+&(\left.-i\hat\partial-\sigma+\mu\gamma^0-g\hat A^3
\frac{\sigma_3}2)
(i\hat\partial-\sigma+\mu\gamma^0+g\hat A^3\frac
{\sigma_3}2)\right ].
\label{kp3_12}
\end{eqnarray}
The operator under the first ${\rm det}$-symbol in (\ref{kp3_12})
acts only in the flavor, coordinate and spinor
spaces, whereas the operator under the second
${\rm det}$-symbol acts in the two-dimensional color subspace,
corresponding to the residual $SU_c(2)$ symmetry of the vacuum, too.
In (\ref{kp3_12}) $\sigma_3={\rm diag}(1,-1)$ is the matrix in the
two-dimensional color space.

\section{Chromomagnetic Catalysis Effect; the case $\mu=0,H\ne 0$ }

The primary goal of the investigations in the present Section is to
clarify the genuine role of the external chromomagnetic field in
dynamical symmetry breaking. In particular, we bring special
attention to the CSC generation. It is well-known that CSC
is induced at sufficiently high values of the chemical potential
\cite{son}-\cite{alf2}. In order to exclude its influence, 
we put here $\mu=0$  and 
study the phase structure of the NJL model (1) at nonzero $H$.

First of all let us study the $H=0$ case. 
Putting $\mu$ and $A_\nu$ equal to zero
and taking into account the general formula $\det O=
\exp({\rm tr}\ln O)$ it is straihtforwardly possible
to perform the calculations of the determinants in (\ref{kp3_12}).
As a result, we have
\ba
\label{eq.11}
V_0(\si,\Del,\Del^\ast)&=&\frac{3\sigma^2}{2G_1}+\f
{3|\Del|^2}{G_2}
-8\int \f{d^3k}{(2\pi)^3}\sqrt{\si^2+4|\Del|^2+k^2}-\nn
&-&4\int \f{d^3k}{(2\pi)^3}\sqrt{\si^2+k^2}.
\ea
This expression has ultraviolet divergences. Hence, we need to
regularize it by cutting off the range of integration:
$|\vec k|\leq\La$. As a result of integrations in the
obtained relation, one can find instead of (\ref{eq.11}) the
following regularized expression:
\be
v_0(x,y)=\f{3A}2 x^2+By^2-\f12\sqrt{1+x^2}-\f{x^2}4 F(x)
-\sqrt{1+x^2+y^2}-\f{x^2+y^2}2 F(\sqrt{x^2+y^2}),
\label{eq.14}
\ee
where the new notations are used:
\[
x=\f{|\si|}\La,~~~ y=\f{2|\Del|}\La,~~~ A=\f{\pi^2}{G_1\La^2},~~~
B=\f{3\pi^2}{4G_2\La^2},~~~  V_0(\sigma,\Delta,\Delta^\ast)=
\f{\La^4}{\pi^2}v_0(x,y),
\]
\be
F(x)=\sqrt{1+x^2}-x^2\ln\f{1+\sqrt{1+x^2}}x.
\label{eq.13}
\ee
There are four different types of stationary points for the 
function (\ref{eq.14}).

Type I point: $(0,0)$. 
It exists for all values of parameters $A,B\geq 0$.

Type II point: $(x_0,0)$. 
It exists for $0\leq A\leq 1$. 

Type III point: $(0,y_0)$.  
It exists for $0\leq B\leq 1$.

Type IV point: $(\ti x_0,\ti y_0)$. It is possible to show that this
solution of the stationarity equations
exists in the region $\om$ of the $(A,B)$ plane, where
\be
\om=\{(A,B)~:~B\geq 0,~B\leq A,~3A-2B\leq 1~\}.
\label{eq.19}
\ee

Let us denote by $v_1,v_2,v_3,v_4$ the values of the potential
(\ref{eq.14}) at the stationary points of type I,II,III,IV,
correspondingly.
In order to find the global minimum point (GMP) of the potential,
we should compare the quantities $v_1,...,v_4$ and select the minimal
one
for each fixed value of parameters $A,B$. As a result of such 
comparisions one can obtain the phase portrait of the 
model at $H=0$, which is presented in Fig. 1. This figure shows the 
$(A,B)$-plane, which is divided into four regions (phases).
These regions are denoted similarly to the stationary points, at
which
the GMP of the potential occurs. So, in the 
region I the GMP is the stationary point of type I, and there is the
totally symmetric phase of the theory.
In the region II GMP corresponds to the $\langle\bar qq\rangle$$\ne
0$,  $\langle qq\rangle$$=0$, hence it is the CSB phase.
The region III is the pure CSC phase, since for all the points from
it
only the diquark condensate is nonzero. Finally, in the region IV,
which is the same as the domain $\omega$ (\ref{eq.19}), the mixed
phase of the theory occurs, since in this case both condensates
are nonzero: $\langle qq\rangle$$\ne 0$, $\langle\bar qq\rangle$$\ne
0$.

We should also note that in the paper \cite{van} the possibility
for CSC at $\mu=0$ was discussed in the 
framework of random matrix models at $H=0.$
Using general symmetry arguments, there a strong constraint on the 
coupling constants, at which the CSC is
forbidden, was obtained. In terms of the NJL model (1) this
constraint
means that at 
$B>A$ the existence of CSC is prohibited. Just the same result
follows from our investigations (see Fig. 1).

Now let us study the influence of a nonzero external chromomagnetic
field with vector-potential (\ref{x7}) and at $\mu=0$ on the
phase
structure of the model (1). In this case one can show from
(\ref{kp3_12}) that (for details see \cite{vdov,e1,1,2}):
\ba
\label{eq.24}
V_{H}(\si,\Del,\Del^*)=\frac{3\sigma^2}{2G_1}+\f
{3|\Del|^2}{G_2}&+&\f{gH}{4\pi^2}
\int_{0}^{\infty} \f{ds}{s^2} \exp
(-s(\si^2+4|\Del|^2))~\coth(gHs/2)\nn
&-&4\int \f{d^3k}{(2\pi)^3}\sqrt{\si^2+k^2}.
\ea
The potential (\ref{eq.24}) is an ultraviolet divergent quantity. 
After regularization,
it can be represented similar to the zero external field case, in the 
following form:
\be
\label{eq.30}
v_h(x,y)=v_0(x,y)-\f{h^2}{2}\Bigl\{
\zeta '(-1,z)-\f12[z^2-z]\ln z+\f{z^2}4\Bigr\},
\ee
where we have used the same notations as in (\ref{eq.14}),
(\ref{eq.13}) as well as
the new ones:
\be
\label{eq.29}
V_H(\si,\Del,\Del^*)=\f {\La^4}{\pi^2}v_h(x,y),~~~~h=\f {gH}{\La^2},
~~~~z=\f{x^2+y^2}h.
\ee
Besides, in 
(\ref{eq.30}) $\zeta'(-1,x)$$=d\zeta(\nu,x)/d\nu|_{\nu=-1}$,
where $\zeta (\nu,x)$ -- is the generalized
Riemann zeta-function.

Numerical and analytical investigations of the potential
(\ref{eq.30}) result in the phase portrait of the model (1) at 
nonzero external field, depicted in Fig. 2 in terms of $A,B$.
First of all one should note that the symmetric phase is absent,
even for arbitrary small values of $H$,$G_1$,$G_2$ (large values of
$A,B$). This is the so called chromomagnetic catalysis effect
of dynamical symmetry breaking. Depending on the relation between
$A$ and $B$, the external field (\ref{x7}) can induce CSB or CSC.
The boundary between pure CSC and mixed phases is the line 
$3A-2B=1$. The boundary between IV and II phases is an $h$-dependent
curve, which is depicted on the Fig. 2 for several values of $h$. 
The left and right boundaries of the region IV asymptotically
coincide at $A,B\to\infty$.
It is necessary to note that the mixed phase IV for arbitrary fixed
$h$ is arranged inside the region $\Om=\{(A,B):0<3A-2B<1\}$.
Obviuosly, $\om\subset\Om$, i.e. under the influence of $H$ both
CSC and mixed phases are spread. Moreover, it is possible to show
that
for an arbitrary fixed point $(A,B)\in\Om$ there is the value
$H_c(A,B)$ of the external chromomagnetic field, such that at
$H>H_c(A,B)$ the point $(A,B)$ lies inside the phase IV.

It can easily be seen from our investigations, that the general
constraint on coupling constants, 
which forbids the CSC and is valid
at $H=0$ (see \cite{van}), is modified at $H\ne 0$.
Indeed, at $H=0$ the CSC is no longer generated at $B>A$ in the
framework of 
model (1). However, at $H\ne 0$ it is forbidden at $B>3A/2$ only.
This stronger restriction is based on the ability 
of the external chromomagnetic field to induce the CSC.

Finally, we should note that, as it was shown in \cite{zhuk,1},
the non-abelian chromomagnetic fields, similar to the abelian ones
of the type (\ref{x7}), are a good catalysts of CSB or CSC.
As in the case of the  ordinary magnetic catalysis effect, 
both the CSB or CSC generation by an external chromomagnetic field
are due to the dimensional reduction mechanism. Notice further that,
if chiral or color symmetry breaking is induced by some types of
external chromomagnetic field at $A,B\gg 1$, then there exists a
critical
temperature $T_c\sim \sqrt{gH}\exp (-C/gH)$ (with $C$ some parameter
depending on the coupling constant), at which the initial symmetry of 
the model is restored (see \cite{zhuk,1}).

\section{the case $\mu\ne 0,H\ne 0$}

In the present Section we consider the influence of nonzero 
values of the chemical potential and external chromomagnetic field on
the
competition between $\langle qq\rangle$- and $\langle\bar qq\rangle$
condensate generations. In this case it is possible to get from 
(\ref{eq.8}) the expression for the effective potential 
(see the paper \cite{2} as well):
\begin{eqnarray} &&
V_{H\mu}(\sigma,\Delta)=N_c\left(\frac{\sigma^2}{2G_{1}}+
\frac{|\Delta^2|}{G_2}
 \right)
- 2N_f\int\frac{d^3p}{(2\pi)^3}(N_c-2)\biggl\{E_p+\theta(\mu-E_p)
\biggr\}-\nonumber\\
   &-&\frac{N_fgH}{8\pi^2}\sum_{n=0}^\infty
  dp_3\alpha_n\biggl\{\sqrt{(\varepsilon_n-\mu)^2+4|\Delta|^2}
  +\sqrt{(\varepsilon_n+\mu)^2+4|\Delta|^2}\biggr\},
\label{eq.20}
\end{eqnarray}
where $E_p=\sqrt{\bar p^2+\sigma^2},$
$\varepsilon_n=\sqrt{gHn+p_3^2+\sigma^2}$
and $\alpha_n=2-\delta_{0n}$.
For convenience, relation (\ref{eq.20}) is written
in terms of $N_f$ and $N_c$ even though in the following we will be
concerned only with $N_f=2$ and $N_c=3$.

{\bf Regularization}.
First of all, let us subtract from (\ref{eq.20}) an infinite
constant in order that the effective potential obeys the constraint
$V_{H\mu}(0,0)=0$.
After this subtraction the effective potential still
remains UV divergent. This divergency could evidently be removed by
introducing a simple momentum cutoff $|\bar p|<\Lambda$. 
Instead of doing this, we find it convenient to use another
regularization procedure. To this end, let us recall that all UV
divergent contributions to the subtracted
potential $V_{H\mu}(\sigma,\Delta)-V_{H\mu}(0,0)$
are proportional to powers of meson and/or diquark fields
$\sigma$, $\Delta$.
So, one can insert some momentum-dependent form factors in
front of composite $\sigma$--and $\Delta$--fields in order to
regularize the UV behaviour of integrals and sums.

It is clear by now that we are going to study the effects of an
external chromomagnetic condensate field in the framework of the 
NJL-type model (1), which in addition to two independent 
coupling constants $G_1,G_2$ includes regularizing meson (diquark)
form factors. Of course, it
would be a very hard task
to study the competition of DSB and CSC for arbitrary values of
coupling constants $G_1,G_2$ and any form factors. 
Thus, in order to restrict this arbitrariness and to be able to
compare our results (at least roughly) with other approaches, 
we find it convenient to investigate  the phase structures
of the model (1) at $H=0$ and $H\ne 0$ only for some fixed values of
$G_1,G_2$ and some simple expressions for 
meson/diquark form factors. We expect that qualitatively the
obtained (integrated) results do not depend significantly on the
chosen regularization procedures, including the momentum cutoff one.

Let us choose the form factors
\footnote{The application of the smooth meson
form factors (\ref{ff}) leads in a natural
way to a suppression of higher Landau levels which is of particular
use here. Hence, this regularization scheme is particularly suitable
for the manifestation of the (chromo)magnetic catalysis 
effect of dynamical symmetry breaking.
Indeed, the (chromo)magnetic catalysis effect and 
the underlying mechanism of dimensional reduction are closely related
to the infrared dominance of the lowest Landau level with $n=0$ \cite
{gus1,zhuk}.} 
\begin{equation}
\phi=\frac{\Lambda^4}{(\Lambda^2+\vec p^{\kern2pt2})^2},\qquad
\phi_n=\frac
{\Lambda^4}{(\Lambda^2+p_3^2+gHn)^2},
\label{ff}
\end{equation}
which have to  be included in  the energy
spectra by a corresponding multiplication of the $\sigma-,\Delta-$
fields:
\begin{equation}
E_p^r=\sqrt {\vec p^{\kern2pt2}+\phi^2 \sigma^2}, \qquad
\varepsilon_n^r=\sqrt {gHn+p_3^2+\phi_n^2 \sigma^2},
\qquad |\Delta^2|\rightarrow \phi_n^2|\Delta^2|.
\label{22}
\end{equation}
Let us denote the regularized expression  for the thermodynamic
potential as $V^r_{H\mu}(\sigma,\Delta)$.
Notice that at $H=0$ it formally coincides with the corresponding
expression of Ref.\cite{berg} 
obtained for an NJL type model with instanton-induced four-fermion 
interactions and taken at zero temperature. 
In particular, by a suitable choice of coupling
constants $G_1,G_2$,
we will later ``normalize'' our phase portraits for $H=0$ to the
curves
of this paper in order to illustrate the influence of a nonvanishing 
chromomagnetic field. 
\footnote{It is necessary to 
underline that in our case the meson/diquark form factors (\ref{ff})
mimic solutions of the BS-equation for some non-local
four-fermion interaction arising from the one-gluon exchange
approach to QCD. Contrary to this, the instanton-like form factor
used in \cite{berg} has another physical nature. It appears as  quark
zero
mode wave function in the presence of instantons \cite{rapp}.}
Despite the $\Lambda$-modification,
the expression for $V^r_{H\mu}(\sigma,\Delta)$ contains yet
UV-divergent integrals.
However, as it was pointed out above, we
shall numerically study the subtracted effective
potential, i.e. the quantity
$V^r_{H\mu}(\sigma,\Delta)-$$V^r_{H\mu}(0,0)$,
which has no divergences.

{\bf Numerical discussions}.
Recall that we have chosen the form factors as in
(\ref{ff}) in order to roughly normalize our numerical
calculations
at $H\ne 0$ on the results obtained at $H=0$ in \cite{berg}.
Comparing the effective potential $V^r_{H\mu}(\sigma,\Delta)$ at
$gH=0$ with the
corresponding one from ref. \cite{berg}
(denoting their respective diquark field and coupling constants by a
tilde), we see that 
these quantities coincide if $2\Delta=\tilde \Delta$,
$G_1=2N_c\tilde G_1$ and 
$G_2=N_c \tilde G_2$. 
Using further 
their numerical ratio of coupling constants, 
we get  in our case the following relation
\begin{equation}
  G_2=3 G_1/8.
\label{gg}
\end{equation}
Now, let us perform the numerical investigation of the global minimum
point (GMP) of the potential $V^r_{H\mu}(\sigma,\Delta)$ for form
factors and values
of coupling constants as given by (\ref{ff}) and (\ref{gg}),
respectively. We use three different values of cutoff:
$\Lambda=$0.6 GeV, 0.8 GeV, 1 GeV. Since the physics should not
depend on $\Lambda$, for each value of $\Lambda$
the corresponding value of $G_1$ is selected from the requirement
that the GMP of the function 
$V^r_{H\mu}(\sigma,\Delta)$ at $\mu=H=0$ is at the
point $\sigma = 0.4 $ GeV, $ \Delta=0$ in agreement with
phenomenological results and \cite{berg}. (Then, the value of $G_2$
is fixed by the relation (\ref{gg}).)
For example, $G_1\Lambda^2=2 N_\up{c}6.47$ at $\Lambda=0.8$ GeV,
$G_1\Lambda^2=2 N_\up{c}6.16$ at $\Lambda=1$ GeV etc.

First of all, it should be remarked that,
as in paper \cite{berg} at $gH=0$, a
mixed phase of the model
was not found for $H\ne 0$, i.e. for a wide range of parameters
$\mu,H$ we
did not find a global minimum point of the potential (\ref{eq.20}),
 at which $\sigma\ne 0$, $\Delta\ne 0$. Since in the 
case under consideration the relation  (\ref{gg}) corresponds
to $B=2A$, this result does not contradict to the 
conclusion of the previous Section.
(Recall, that at $\mu=0$ the diquark condensation is prohibited
in the region $B>3A/2$.) 

The results of
our numerical investigations of the GMP of the effective potential
are
graphically represented  in the Fig. 3. For each value of the cutoff
$\Lambda$ the phase portrait of the model consists of two phases II
and III. The boundary between the two phases is practically 
$\Lambda$-independent and represents a first order phase
transition curve. It is necessary to note also that for 
each of the above mentioned values of $\Lambda$ and for fixed
value of $gH$ there is a critical chemical potential
$\mu_c(H)$, at which the GMP is transformed from a point of type III
to a symmetric point of type I. However, this phase transition is 
a significantly $\Lambda$-dependent one. Indeed, even in the simplest
case with $H=0$ we have 
 $\mu_c(0)=1$ GeV at $\Lambda=0.6$ GeV,
$\mu_c(0)=1.3$ GeV at $\Lambda=0.8$ GeV,
$\mu_c(0)=1.65$ GeV at $\Lambda=1$ GeV. Hence, in the framework of
the NJL model (1) such a phase transition 
is just an artefact of the regularization procedure, which agrees
with
 the QCD results of \cite{son,raj} that CSC can exist at enormously
 high
values of chemical potential $\mu\gtrsim 10^8$. By this reason, it is
not indicated on the Fig. 3.

On the phase portrait of Fig. 3  $\mu$ and $gH$ are free
parameters. However, as was emphasized in the Introduction,
the external chromomagnetic field mimics the 
gluon condensate, so the value of $gH$ is some definite 
quantity. Here we should note that
recent investigations yield the following value of 
the QCD gluon condensate at $T=\mu=0$:
$gH\approx 0.6$ GeV$^2$ \cite{glue}. In the paper \cite{saito} it was
shown in the framework of
a quark-meson model that at ordinary nuclear density
$\rho_0$ the gluon condensate decreases by no more than six percent,
compared with its value at zero density. At densities $3\rho_0$ the
value of  $\langle FF\rangle$ decreases by fifteen percent. This
means that for values
of the chemical potential $\mu <1$ GeV the gluon condensate is 
a slowly decreasing function vs $\mu$. Taking in mind this 
circumstance, one can draw two important  conclusions from our
numerical analysis. Firstly, at $H=0$ and $\mu=0.4$ GeV there 
should exist the 
CSC phase (see \cite{berg}). However, if the real gluon condensate 
$gH\approx 0.5$ GeV$^2$ is taken into account at $\mu=0.4$ GeV,
and assuming that our results would remain valid also for more
realistic condensate fields, this would seemingly render it more 
difficult to observe the CSC phase in heavy ion-ion experiments. 
 Secondly, let us discuss some quantitative 
characteristics of the CSC phase at $\mu=0.8$ GeV. As it follows
from our numerical analysis, at
$\mu=0.8$ GeV, $gH=0$,  the GMP of the effective potential
corresponds to the CSC phase with a stable diquark condensate
$\Delta\approx 0.1$ GeV. However, assuming that 
the value of the gluon condensate $gH\approx 0.4$
GeV$^2$ would hold for the above nonvanishing chemical
potential,
one would get a value of the  diquark condensate $\Delta\gtrsim 0.2$
GeV, which is significantly larger in magnitude, than at
$gH=0$ (this estimate was obtained for the case $\Lambda=1$ GeV).

As a general conclusion, we see that taking into account an
external chromomagnetic field
at least in the form as considered in the model above,
might, in principle, lead to  remarkable
qualitative and quantitative changes in the 
picture of the
diquark condensate formation, obtained in the framework of 
NJL models at $H=0$.

\section{Summary and conclusions}

In the present talk the ability of external
chromomagnetic fields to induce dynamical symmetry breaking (DSB) of
chiral and color symmetry
was studied in the framework of the extended NJL model (1) with
attractive quark interactions
in  $qq$- and $\bar qq$-channels. Particular attention was paid
to the CSC generation. First of all, in order to understand the
genuine role of an external chromomagnetic field in the CSB or CSC
phenomenon, we have removed the chemical potential from our
consideration.
The numerical analysis shows in this case (see Fig. 2) that even at 
sufficiently small values of coupling constants the
external chromomagnetic field catalyzes the DSB of chiral and color
symmetries (the chromomagnetic catalysis phenomenon).
This effect
is accompanied by an effective lowering of dimensionality in strong
chromomagnetic fields, where the number of reduced units of
dimensions depends on the concrete type of the field ---
a conclusion already made in the case of the CSB \cite{zhuk}.
As was shown in \cite{zhuk,1}, the phenomenon of $qq$-
as well as $\bar qq$-condensation
does exist for various (non-)abelian chromomagnetic field
configurations in the weak coupling limit.

The possibility for vacuum CSC at $\mu=0$ was also studied in the
framework
of random matrix models on the basis of general symmetry arguments
\cite{van}. There it was found a constraint on the coupling
constants in $qq$- and $\bar qq$-channels, at which the CSC
is forbidden.  We have, in particular, shown that the
external chromomagnetic
field modifies this constraint and reduces the region of
coupling constants, in which the CSC cannot occur (see Section III).

Then, we have considered a more realistic case with nonzero 
chemical potential as well as with physically meaningful values
of coupling constants (Section IV). It is well-known that in this
case at rather moderate values of chemical potential ($\mu\gtrsim
0.3$ GeV) the new CSC phase of QCD is predicted
\cite{rapp}-\cite{alf2}. However, in these papers such
nonperturbative
feature of the real QCD vacuum as the nonzero gluon chromomagnetic
condensate was not taken into account. In the present analysis
in the framework of NJL model (1), the gluon condensate is
simulated as an external chromomagnetic field,
i.e. $\langle FF\rangle\equiv$$2H^2$.
Our numerical calculations show that for real values of the 
gluon condensate the CSC phase, in contrast with results of
\cite{rapp}-\cite{alf2}, cannot appear for low chemical
potentials 
0.3 GeV$<\mu<$0.6 GeV. 
At larger values of $\mu$
the gluon condensate significantly modifies the value of the
diquark condensate obtained at $H=0$.

Thus, the main conclusion of our investigations is that the inclusion
of an
external chromomagnetic field might significantly change the
picture of CSC formation, obtained at $H=0$.

\section*{Acknowledgements}
We wish to thank V.P. Gusynin, V.A. Miransky and Y. Nambu 
for fruitful discussions. D.E. gratefully acknowledges the support
provided to him by the Ministry of Education and Science and
Technology of Japan (Monkasho).
This work is supported in part by DFG-Project 436 RUS 113/477/4.

\vspace{1cm}

\begin{center}
{\bf FIGURE CAPTIONS}
\end{center}

\noindent Fig.1.
The phase $(A,B)$-portrait of the model at $H=0$.

\noindent Fig.2. The phase $(A,B)$-portrait of the model for
several values of $h=gH/\Lambda^2$.

\noindent Fig.3. The phase $(gH,\mu)$-portrait of the model for
several values of $\Lambda$, which are indicated in GeV's.

\vspace{1cm}

\begin{figure}\centering
\epsfxsize=80mm\epsfbox{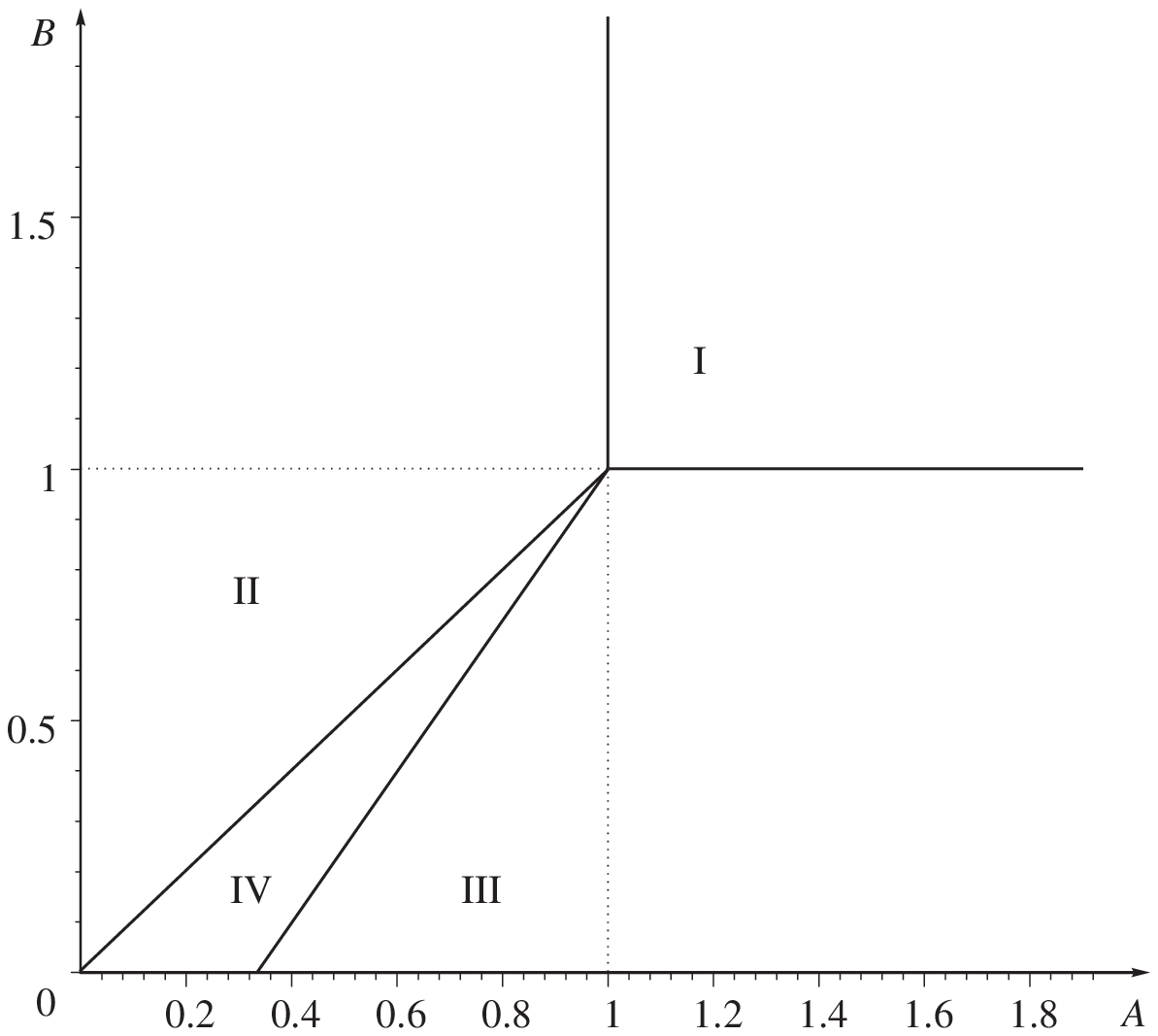}
\caption{}
\end{figure}

\begin{figure}\centering
\epsfxsize=80mm\epsfbox{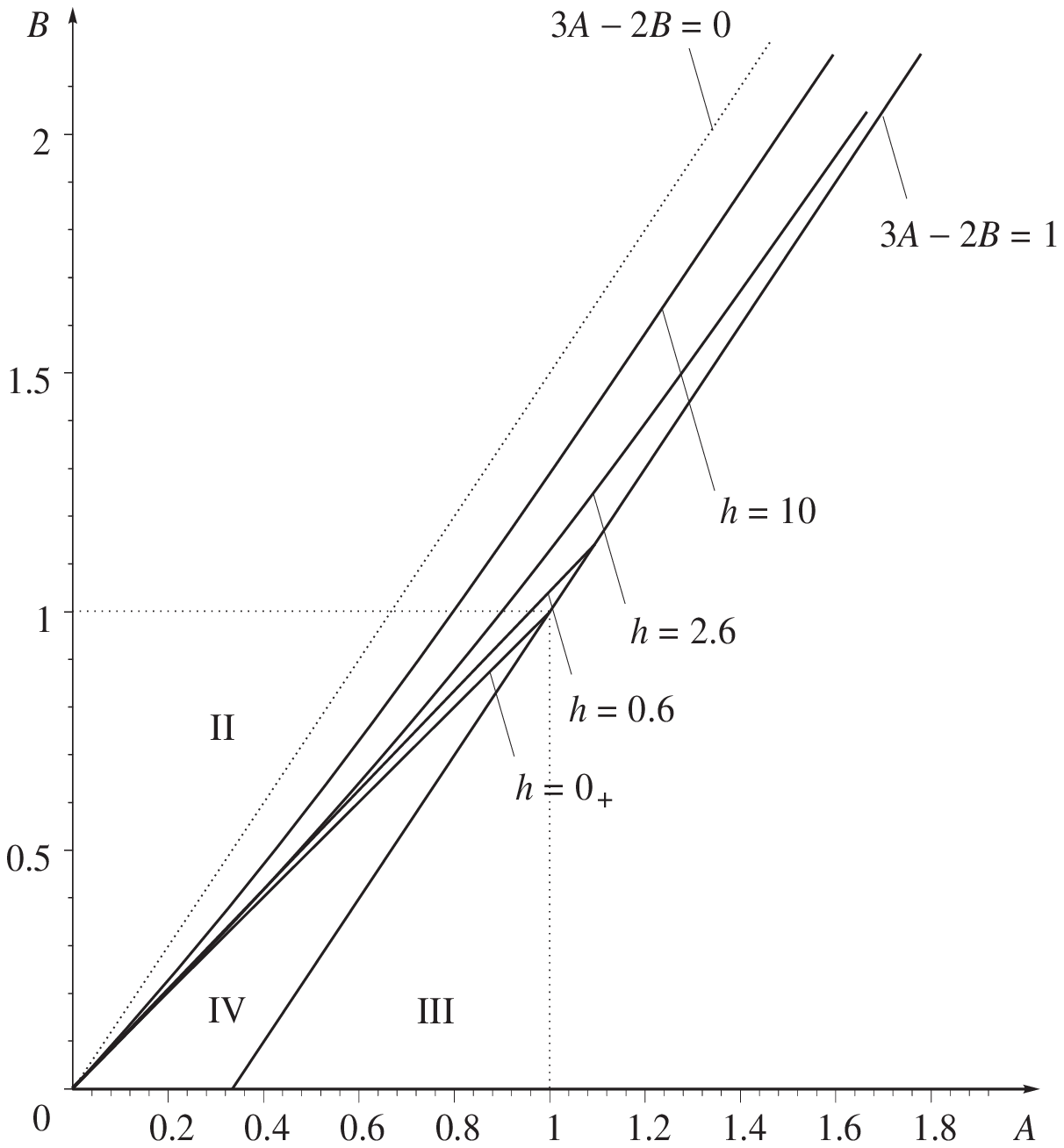}
\caption{}
\end{figure}

\begin{figure}\centering
\epsfxsize=80mm\epsfbox{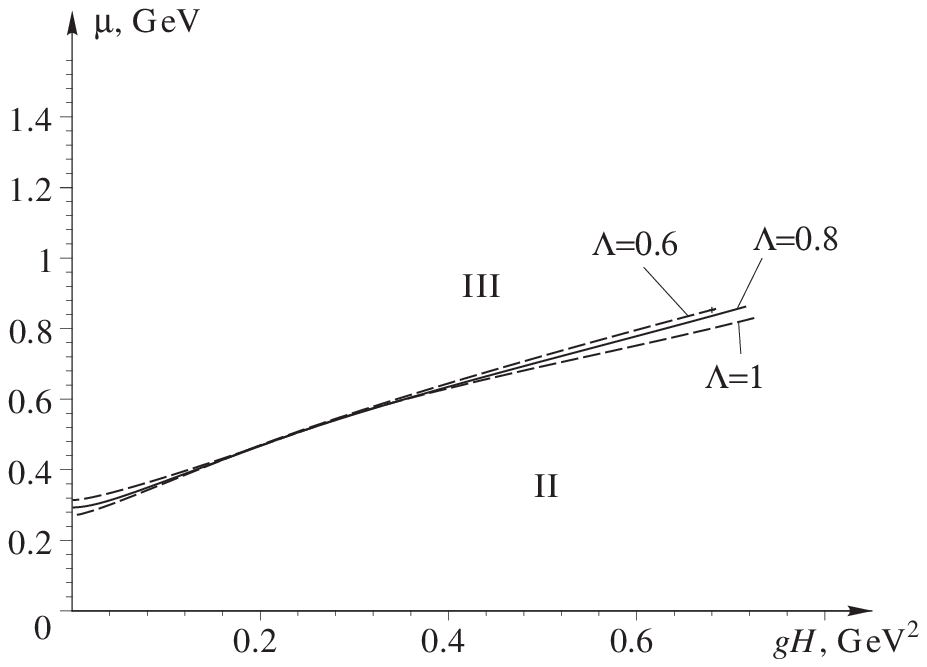}
\caption{}
\end{figure}

\end{document}